\documentstyle[12pt]{article}
\def\beq{\begin{eqnarray}}    
\def\eeq{\end{eqnarray}}      
\def\tr{\,\mbox{tr}\,}                  
\def\Tr{\,\mbox{Tr}\,}                  

\def\det{\rm det }

\def\BV{Barvinsky-Vilkovisky}		

\def\al{\alpha}
\def\be{\beta}
\def\ch{\chi}
\def\ga{\gamma}
\def\de{\delta}

\def\la{\lambda}
\def\na{\nabla}

\def\si{\sigma}
\def\om{\omega}

\def\up{\upsilon}
\def\Ga{\Gamma}

\def\Om{\Omega}

\def\CO{{\cal{O}}}
\def\sign{{\rm sign}}
\def\inbar{\,\vrule height1.5ex width.4pt depth0pt}
\def\IC{\relax\hbox{$\inbar\kern-.3em{\rm C}$}}

\begin{document}

\hfill DFTUZ 96-15,$\,\,\,\,\,\,$ hep-th/9610006

\hfill September, 1996
\vskip 5mm

\begin{center}

\setcounter{page}1
\renewcommand{\thefootnote}{\arabic{footnote}}
\setcounter{footnote}0

{\Large \bf Some remarks on high derivative quantum gravity}

\vskip 8mm
{\bf M. Asorey$^a$, J.L. L\'opez$^a$ and
I.L. Shapiro$^{a,b}$ }\footnote{ On leave from Tomsk State Pedagogical
Institute, Tomsk, 634041, Russia}

\vskip 3mm

{ a) Departamento de F\'{\i}sica Te\'orica.  Universidad de Zaragoza
\\ 50009, Zaragoza, Spain}

\vskip 3mm

{ b) Departamento de F\'{\i}sica
-- ICE, Universidade Federal de Juiz de Fora
\\ 33036-330, Juiz de Fora -- MG, Brazil.}

\end{center}

\vskip 8mm
\begin{center}
{\bf  Abstract}
\end{center}
\noindent
\baselineskip 14pt
We analyze  the perturbative implications of
the most general high derivative approach to quantum gravity based on
a diffeomorphism invariant local action. In particular, we consider
the super-renormalizable case with a large  number of metric
derivatives in the action. The structure of ultraviolet
divergences is analyzed in some detail. We show that they are
independent of the gauge fixing
condition and the choice of field reparametrization. The
cosmological  counterterm is shown to vanish under certain
parameter conditions. We elaborate on the unitarity
problem of high derivative approaches and the distribution of masses
of unphysical ghosts. We also discuss the properties of the low
energy regime and explore the possibility of having a  multi-scale
gravity with different scaling regimes compatible  with Einstein
gravity at low energies. Finally, we show that the  ultraviolet
scaling of matter theories is not affected by the quantum
corrections of high derivative gravity. As a consequence, asymptotic
freedom is stable under those quantum gravity corrections.

\vskip 12mm

\noindent
{\large \bf Introduction}
\vskip 2mm
\baselineskip 16pt
The formulation of a consistent theory of Quantum Gravity
is still one of the major challenges in theoretical physics.
One of the main fundamental problems is that there is no experimental
evidence of any Quantum Gravity effect \cite{ish}, whereas
  the classical theory covers from
cosmology to current precision tests with great success.
 The situation is considerably
more dramatic that for Quantum
Electrodynamics  at the end of the forties. The reason being that in the
last case, although
the theory was not completely consistent, there was experimental
evidence on the existence of relativistic quantum effects associated to
electron dynamics. Nature provided  confidence on the co-existence
of special relativity and
quantum mechanics although field theory was not yet at sight.
An analogous conviction does not exist in
Quantum Gravity.
On the other hand since  the quantum effects can
 only, in principle, be  observed at the Plank scale which
lies too far from  current experiments, the lack of
experimental data is not unnatural. Consequently,
any theoretical model of Quantum Gravity has to be
necessarily based on  theoretical principles
without phenomenological constraints.

 On spite of this  freedom   of the theory
of Quantum Gravity,
the construction of  consistent  models meets very
serious difficulties. First, the
quantum theory based on the Einstein-Hilbert action is
non-renormalizable \cite{hove,dene}.
The radiative corrections to the effective action
contain divergences which involve a number of derivatives of the
metric in the counterterms which is increasing with the number of
loops. Indeed, one can use the general theorems on  covariant
renormalization (see \cite{voty}
and references therin)
to show that all these counterterms are general covariant local
expressions. Thus, it is possible to remove all the divergences starting
from the Einstein theory by adding an infinite tower  of  possible
higher derivative terms,  regarded as the perturbations.
In that case
the definition of the renormalized theory requires, in general,
the introduction of an infinite number of renormalization
conditions, unless some parameter reduction mechanism appears
\cite{acs}.
 Recently, it was pointed out \cite{don} that some large distances
 effects remain, however, independent of the renormalization conditions
involving higher order terms of the effective action.
In particular this occurs
for the quantum corrections to the gravitational Newtonian potential.
However,
the analysis of other effects like the scaling dependence of the
cosmological
and gravitational constants, depends on the concrete formulation of
the quantum theory.

One can always
imagine that the extremely high energy UV regime is described by qualitatively
different theory like  string theory, free of  renormalizability problems,
 which will provide a natural reduction parameter
scheme at intermediate scales. Having this as perspective, field
theoretical approaches to quantum gravity should be considered as
effective field theories. However the predictions of string theory
quantum gravity effects at the intermediate energies still remain unveiled.
On the other hand, the string effective action is well defined only on shell.
Off shell continuations are not uniquely defined because of
reparametrization invariance \cite{zwe}--\cite{bbhr}.
For such a reason, we will consider
the most general effective action\footnote{We
do not consider the dilaton and the antisymmetric fields
for the sake of simplicity throughout this paper.
The inclusion of such fields does not introduce  qualitative changes  in
 the results.} which can be generated by string
theory \cite{gsw}
\beq
S_{\rm eff}(g_{\mu\nu}) = \lim_{N\rightarrow \infty}
\sum_{n=0}^{N+2}
\al_{2n}\;m^{4-2n}\;\int d^4x \sqrt{-g}\; \CO_{2n}(\partial_\la g_{\mu\nu}),
\label{1}
\eeq
where   $ m^2$
is the only dimensional parameter of our fundamental theory
(in string theory it is   the string
tension $1 / \al'$) while  $\CO_{2n} (\partial_\la g_{\mu\nu})$
denotes the general covariant scalar terms
containing $2n$ derivatives of the metric $g_{\mu\nu}$ and
the constants $\al_n$
are dimensionless couplings.
Although theories with higher derivatives like (\ref{1}) are  in general
non-unitary at the quantum level, string theory is both unitary and
renormalizable. In particular,
one can always choose a
special parametrization without unphysical  ghosts
\cite{zwe,dere}, \cite{den4}, although from a pure string point of view
there are no means to distinguish this special parametrization
of the metric except for the absence of
ghosts \cite{tse} (see also \cite{ovrut}).
 However
if one  considers an approximation to the effective theory and makes a
truncation
of the series (\ref{1}),  the  unitarity problem   reappears.
The problem   shows up even at the classical level where, because
of the  existence of the unphysical ghosts with
negative energy, it leads to  classical instabilities \cite{ste78}.
The quantum unitarity problem has been considered in great detail in
the particular case of (four derivative) $R^2$-gravity
\cite{utdw,dene}, \cite{ste}-\cite{bush}.
In that case, the ultraviolet behaviour of  the
propagators and vertices leads to a renormalizable theory
\cite{ste,voty}. Unfortunately the particle content of this theory
contains, besides the massless graviton,
a spin-$2$ massive unphysical ghost which violates unitarity
\cite{ste,ste78}. Despite a lot of interesting attempts to solve the
unitarity problem
in  $R^2$-gravity \cite{sast}-- \cite{dgo} it turns out that the
massive ghost is not removed by radiative corrections, and
therefore  high derivative theory
can not be considered as the fundamental theory of quantum gravity
\footnote{It is a remarkable fact that all attempts to derive
a Field Theory of Gravity from non-commutative geometry lead to
$R^2$-gravity}. The appearance of the massive ghosts is the price
to pay for   renormalizability. Their contributions are
essential to reduce the dimension of  counterterms. However, since
the masses of the ghosts are of the order of the Planck mass, the
fourth derivative quantum gravity can be successfully used as an
effective theory at the energies below this scale, where the
unphysical ghosts are not generated \cite{frts}-\cite{bush}.

 The aim  of the present paper is to extend those results for the
general case of high derivative gravity. We will truncate
 the effective string theory action $S_{\rm eff}$
 taking a finite value for  $N$.  If $N> 0$ the corresponding theory
  contains more that four derivatives of the
metric  and it  becomes
super-renormalizable. This provides a natural framework to study
the possibility of different scaling regimes  compatible
with Einstein gravity at low energies and  different
scenarios for the ultraviolet behaviour of the cosmological
constant.

The organization of the paper is as follows. In section 2 we consider the
 quantization of the general higher derivative theory  and show its
super-renormalizability. In section 3 the general structure of ultraviolet
divergences is analyzed and the cosmological counterterm is
explicitly calculated for the general case.
We also study the gauge fixing independence
of  the counterterms.
The low energy regime of the theory is examined and
shown to be equal to that of Einstein gravity in section 4,
where we also discuss the spectrum of massive ghosts, and
 the possibilities of having different scaling regimes.
 In section 5 we study the coupling of high derivative
gravity to general gauge theories and show that their beta
functions are not modified by the gravitational corrections.
Finally, a discussion of the results is developed in  section 6, and
some technical aspects are  postponed to three appendices.

\vskip 6mm

\noindent
{\large \bf 2. Gauge fixing, quantization and power counting.}
\vskip 2mm

The first two terms of the action $S_{\rm eff}$ are the cosmological
term
$$
S^{(0)}_{\rm eff}(g_{\mu\nu})= m^4 \alpha_0 \int d^4x\,\sqrt{g}
$$
 and the Hilbert-Einstein action,
$$
S^{(2)}_{\rm
eff}(g_{\mu\nu})= m^2 \alpha_2 \int d^4x \, \sqrt{g}\,R
$$
The third term  ($n=2$)
 leads to the mentioned fourth derivative gravity
$$
S^{(4)}_{\rm eff}(g_{\mu\nu})= \int d^4x \, \sqrt{g}\,\left(
\alpha^1_4\, R_{\alpha \beta}\,R^{\alpha \beta} + \alpha^2_4\, R^2 \right)
$$
(see \cite{book} for an introduction and more complete references).
Higher order terms ($n>2$) can be expressed in terms of the
Riemann curvature tensor, Ricci tensor, scalar curvature and their covariant
derivatives. For $n=3$ we have  two different types of terms
$R^{(3)}$ and $\na R\cdot\na R$ and for $n=4$  we have
$R^4$, $\,R \na R \cdot \na R$,
$\,\na^2 R \cdot \na^2 R$.   $N$-th order   include terms from
$R^{N+2}_{...}$ and $R^{N-2}_{...}\,\na R^{...}\,\na R^{...}$
to $R_{...}\,{\Box}^N\,R^{...}$. The dots indicate all possible
 contractions of  tensor indices.

Perturbation theory is generated by the standard expansion
around  the flat metric
$$
g_{\mu\nu} = \eta_{\mu\nu} + h_{\mu\nu}.
$$
The    contributions of the $N>0$ terms  to the propagator
can  only come from  terms of second order in the curvature. Terms of third
and higher orders in the curvature  contribute only to  vertices,
because they involve more than two $h_{\mu\nu}$ fields.
Using  Bianchi identities
\beq
\na_\la R_{\mu\nu\al\be} + \na_\be R_{\mu\nu\la\al} +
\na_\al R_{\mu\nu\be\la} = 0
\label{2}
\eeq
 one can easily reduce the terms of second order in the curvature
(with total $\,2n\,$ derivatives) to the form
\beq
\alpha^1_{n}\, R_{\al\be}{\Box}^{n-2}R^{\al\be} +
\alpha^2_{n}\, R\; {\Box}^{n-2}R+
\alpha^3_{n}\, R_{\mu\nu\al\be}{\Box}^{n-2}R^{\mu\nu\al\be}
\label{3}
\eeq
Now one can use (\ref{2}) again as it was done in \cite{dere}
for the $N=1$ (six-derivative) terms to get
\beq
R_{\mu\nu\al\be} {\Box}^{n-2} R^{\mu\nu\al\be} &=&
- \na_\la R_{\mu\nu\al\be} {\Box}^{n-2} \na^\la R^{\mu\nu\al\be}
+ \CO(R^3) +
\na_\mu \Om^\mu \cr
&&\label{4}
\cr
&=& 4 R_{\al\be} {\Box}^{n-2} R^{\al\be} - R {\Box}^{n-2} R +
\CO(R^3) + \na_\mu \Om'^\mu
\eeq
This is indeed similar to the relation which takes place in the
fourth derivative gravity because of the topological nature of
the Gauss-Bonnet term \cite{hove}. Of course, for any $n \neq 2$
the term
\beq
\int d^4x \sqrt{-g}\,
\left( R_{\mu\nu\al\be}\,  {\Box}^{n-2}\,  R^{\mu\nu\al\be} -
4\,  R_{\al\be}\,  {\Box}^{n-2}\,  R^{\al\be} + R\,  {\Box}^{n-2}\,  R
\right)
\label{4j}
\eeq
is not topological and
gives rise to  non-trivial contributions to the vertices.
Since
the $\CO(R^3)$ terms do not contribute to the propagator, the
 relevant terms of order $n>2$ can be written as
\beq
\alpha^1_{n}\,R_{\al\be}\,{\Box}^{n-2}\,R^{\al\be} +
\alpha^2_{n}\,R\;{\Box}^{n-2}\,R                  \label{5}
\eeq
If $\alpha^1_{2N+4}
 \neq 0$ and $\alpha^1_{2N+4} + 3 \alpha^2_{2N+4} \neq 0$,
the highest order
 terms of (\ref{5}) are nondegenerate once we introduce a gauge fixing
and asymptotically  they  behave like
$\CO( k^{2n+4})$ for large momenta. The gauge fixing condition
can be introduced within the standard Faddeev-Popov prescription. In order to
improve the regularity properties of the quantum fields it
is convenient to add a
higher derivative covariant operator
for the longitudinal modes in the gauge
fixing condition \cite{ste,book} (see also  \cite{fs}--\cite{aa} for a
general discussion on gauge
theories)
\beq
S_{\rm gf} = \int d^4x \; \ch_\mu \;C^{\mu\nu} \; \ch_\nu
\label{6i}
\eeq
where
$$
\ch_\mu = \partial_\la h_\mu^\la - \be\, \partial_\mu
h^{\lambda}_\lambda
$$
\beq
C^{\mu\nu}
= - \sum_{n=2}^{N+2}\, \left[ \,\sigma_n\, g^{\mu\nu}\, \Box +
(\ga_n - 1)\,\na^\mu \na^\nu \,\right] \;
\left(\frac{\Box}{m^2}\right)^{n-2}
\label{6}
\eeq
 $\be, \sigma_n, \ga_n$ being dimensionless gauge fixing parameters.
Regardless to the number of derivatives, the propagator of the quantum
metric can be written in terms of the irreducible spin $2,1,0$ projectors
and  some spin-zero transfer operators \cite{ste}. By choosing the
gauge fixing parameters $\be, \sigma_n, \ga_n$  in a special way one can
always remove all the spin-1 states from the spectrum.
The spin-2 states are gauge-fixing independent, just as in the fourth
derivative gravity, and if $\alpha^1_{2N+4} \neq 0$, the propagator
of the spin two states has the asymptotic ultraviolet behaviour
 $\CO( k^{-(2N + 4)} )$.
The same occurs for
 the propagator of  spin-0 states but there the condition reads
$\alpha^1_{2N+4} + 3 \alpha^2_{2N+4} \neq 0$.
Throughout we shall assume   that both
conditions are satisfied and therefore
the propagator of the quantum
metric behaves like $\CO( k^{-(2N + 4)} )$ for
large momenta.
One can absorb the determinant of  the covariance operator
$C^{\mu\nu}$ into the action of Faddeev-Popov ghosts \cite{frts},
\beq
S_{FP}
= \int d^4x\sqrt{-g}\; \left[\,{\bar c}^\la
\; C_{\mu\nu}\,\na^\nu \;(\na^\mu c_{\la}+\na_\la c^{\mu})
- 2 \,\be\,{\bar c}^\la \; C_{\la\nu}\,\na^\nu \;(\na_\mu
c^{\mu})\right] \label{6j}
\eeq
and get that their propagator also behaves
like $\CO( k^{-(2N + 4)})$ in the UV regime.
The extra contribution of the operator $C$ into the ghost
sector has to be  compensated by the corresponding power of
the determinant $\, \det C^{\mu\nu\,}\,$  \cite{frts}
(see also \cite{fs}--\cite{aa} for a general discussion
on gauge theories). In summary, the partition
function is given by
\beq
\int\,[\delta h_{\mu\nu}]\;
[\delta c_\tau ]\; [\delta {\bar c}^\la]\;
\left( \det \,C^{\mu\nu}\right)^{-{1\over 2}} \;
e^{i S\left(g_{\mu\nu}\right) + i S_{\rm gf} + i S_{FP}}
\eeq
With that gauge fixing choice  the propagators
of both metric and ghosts have the same asymptotic
behaviour in the UV limit $\CO(k^{-(2N + 4)})$. Unfortunately
 the interaction vertices pick up all possible number of derivatives
 from zero to $2N + 4$, in such a way that one loop divergences
remain unregularized. Indeed, if we evaluate the superficial degree
of divergency $D$ of an arbitrary
$p$-loop diagram with
$n_{2r}$ vertices with $2r$ derivatives   $\;r = 0, 1,
\cdots, N+2$ we get
\beq
D + d_{\rm ext}
= 4 + 2 N - 2\,N\,p  -
\sum _{r=0}^{N+1}(2\,N + 4 - 2\,r)\;n_{2r}
\label{7}
\eeq
where $d_{\rm ext}$ is the total number of derivatives acting on external
lines. From the power counting identity
(\ref{7}) it follows that  divergent graphs satisfy the
inequality
\beq
d_{ext} \leq 4 + 2\,N - 2\,N\,p
\label{8}
\eeq
which implies that divergences can only appear for  higher loops
 diagrams with $N = 0,1,2$. For $N>2$
divergences appear only in the one-loop diagrams.
For any $N>0$  the   divergent terms involve less
 powers of the curvature  than  those bare action, and moreover,
this power is decreasing with the loop order. We remark that the covariance
of all the counterterms (in an invariant regularization)
is guaranteed by the general theorems \cite{voty}
which can be trivially generalized for diffeomorphism invariant
theories  of the form $S_{\rm eff}$.

Thus,  for any finite $N>0$, the  structure of divergences becomes
 simpler, the
theory  is super-renormalizable. For  $N>2$ the situation is even
better because all the divergences appear at one loop order of perturbation
theory. Now, in any case the structure of one loop divergences is not
changed. This is very similar to what happens in gauge theories where
the method of high covariant derivatives does not smooths the behaviour
of one loop contributions \cite{fs}.

If $N>2$, since the  only
UV divergences appear in one loop diagrams, they can be removed by
 Pauli-Villars determinant regulators using the
 methods first introduced for gauge theories. In such a case we end up with
a completely finite theory (all beta functions vanish) and the corresponding
theory may be considered as a complete regularization of Quantum Gravity.
This is nothing but the  implementation for quantum gravity
of the high derivative method introduced by
Slavnov \cite{sl}.
In such a regularization one can look  for non-perturbative
effects and specially for non-trivial fixed points where the  theory
might be unitary and  renormalizable. If  such fixed points
would exist most of the technical problems mentioned in the
introduction will be overcome.

For any $N>0$ the theory is super-renormalizable, because
the local covariant counterterms have less derivatives than
the classical action and the coefficients of the terms with more derivatives
do not need any kind of infinite renormalization. Thus,  the bare values
of those coefficients can
be kept finite. Their explicit value  can only be fixed by looking at the
 phenomenological implications of the theory or
the predictions of the fundamental theory
which generates our theory as low energy effective theory.
The only sector of the theory
which is subject of infinite renormalization is of the form
\beq
\int d^4 x \sqrt{-g} \;\left\{\,
\alpha^1_{4}\,R_{\al\be} \,R^{\al\be} + \alpha^2_{4}\, R^2 +
\alpha_{2} \,m^2\;R +  \alpha_{0}\,m^4\,\right\}
+ \;(\rm surface\;\; terms)
\label{9}
\eeq
After renormalization  the scaling behaviour of the physical quantities
is governed by a renormalization group equation.
Since the only  parameters which undergo an infinite renormalization
 are those contained in the lower derivative part (\ref{9})
of the action, the scale dependence of the
effective theory $S_{\rm eff}$
is encoded in the beta-functions for the four parameters
$\be_{\alpha^1_4},\; \be_{\alpha^2_4}, \;\be_{\alpha_2},
\;\be_{\alpha_0}$.
We shall focus on the evaluation of  these beta-functions.

Indeed for  large $N$ the Feynman rules are rather involved and
in general even one loop calculations are very difficult.
In the next section we develop a method   to calculate
these beta-functions in one-loop approximation for the most general case,
 and we shall perform an explicit calculation of the cosmological
constant beta function
$\be_\Lambda=\be_{\alpha_0}$.
Our method is essentially based on the techniques introduced
by Barvinsky
and Vilkovisky \cite{bavi}, which can be, in principle, applied for the
calculation of $\be_{\alpha^1_4},\; \be_{\alpha^2_4}, \;\be_{\alpha_2}$
as well.
The only difficulty    is that the manipulation of
algebraic expressions becomes much more involved in those cases.

\vskip 6mm

\noindent
{\large \bf 3. One-loop results.}
\vskip 2mm

Before entering into the calculation of beta-functions, let us show
that they are independent of
the choice of the gauge fixing parameters (\ref{6})
and the parametrization of quantum gravitational fields.
 Local counterterms can be, in principle, gauge and parametrization
dependent, but we will show that these dependence   vanishes for
high derivative gravity. This follows from the fact that for $N>2$
the only UV divergences appear at one loop order and the explicit
relation which exists
between the divergent one-loop
counter-terms associated to two different
sets of the gauge fixing parameters $\sigma=(\sigma_n,\be,\ga_n)$ and
${\bar\sigma} = ({\bar\sigma_n}, {\bar \be}, {\bar \ga}_n)$
\cite{avr,shja} (see also Appendix B for a selfcontained derivation).
\beq
{\Gamma_{\rm div}}({\sigma}) - {\Gamma_{\rm div}}({\bar{\sigma}})
= \int {d^4}x \sqrt {- g} \; \frac{\delta S}{\delta g_{\mu\nu}}\;
\Sigma_{\mu\nu} (g_{\al\be}, {\sigma}, {\bar{\sigma}}),
\label{m1}
\eeq
$\Sigma_{\mu\nu} (g_{\sigma\be}, {\sigma}, {\bar{\sigma}})$ being
 some  local function of metric and gauge fixing parameters.
Power counting tells us that for any choice of the gauge fixing
parameters the divergent counterterms are local expressions
with up to four derivatives of the metric. Thus, the  left hand side
of the identity (\ref{m1}) can only contain such a type of terms.
But the right hand side  which is proportional to classical motion equations
contains terms with $2N+4$ derivatives. Hence the equality can  hold
 if only and only if
$\Sigma_{\mu\nu} (g_{\al\be}, {\sigma}, {\bar{\sigma}})$
is identically zero. One-loop divergences  are, therefore, gauge
independent.
 Since $N>2$ there are not more divergences in higher loops and,
therefore, we have proved that all the beta-functions
$\be_{\alpha^1_4},\; \be_{\alpha^2_4}, \;\be_{\alpha_2},
 \;\be_{\alpha_0}$ do not depend on the gauge  fixing condition.
  In fact, by means of Nielsen identities
\cite{stel}
it can be proved that
even in the case $0 < N \leq 2$ there is no dependence on the
gauge fixing
parameters  in the divergent counter-terms generated by
higher loops.
 The independence on the parametrization
of the quantum metric can be proved in a similar way because two different
reparametrizations lead to divergent contributions which satisfy an
equation similar to (\ref{m1}).

Let us now calculate the one-loop radiative corrections.
The leading ultraviolet term  is given by
\beq
\alpha^1_{2N+4}\,R_{\al\be}\,{\Box}^{N}\,R^{\al\be}
 + \alpha^2_{2N+4}\,R\,{\Box}^{N}\,R
\label{10}
\eeq

We shall assume that $\alpha^1_{2N+4} \neq 0$ and $\alpha^1_{2N+4}
+3\alpha^2_{2N+4} \neq 0$. In the background field method   the metric is
split into a background metric $g_{\mu\nu}$ and quantum metric
$h_{\mu\nu}$,   $$g_{\mu\nu} \rightarrow g'_{\mu\nu} =
g_{\mu\nu}+h_{\mu\nu}. $$
We introduce the background field gauge fixing condition
\beq
S_{\rm gf} = \int d^4x \;\sqrt{-g}\; \ch_\mu \;C^{\mu\nu} \; \ch_\nu
\label{11}
\eeq
where
$$
\ch_\mu = \na_\la h_\mu^\la - \be \;\na_\mu h
$$
\beq
C^{\mu\nu}
= - \frac1\al \;\left( g^{\mu\nu} \,\Box +
\ga \,\na^\mu\, \na^\nu - \na^\nu \,\na^\mu \right) \;
\left(\frac{\Box}{m^2}\right)^{N}                        \label{12}
\eeq
which is a covariant generalization of (\ref{6i})
defined by  replacing ordinary derivatives by
covariant derivatives with respect to the
background metric $g_{\mu\nu}$. The coefficients of the
gauge fixing condition $\al,\be,\ga$ are arbitrary parameters.
Many of the coefficients of (\ref{6}) have been set equal to zero
to simplify the calculations, but the
divergences will not depend on their values.
The one-loop effective action is
\beq
\Gamma^{(1)} = \frac{i}{2} \Tr \ln H_{\mu\nu,\al\sigma} -
i\, \Tr \ln M_\al^\sigma - \frac{i}{2} \Tr \ln C^{\mu\nu}
\label{13}
\eeq
where
\beq
H_{\mu\nu,\al\sigma} =\left.
\frac{\de^2 S_{\rm eff}}{\de h_{\al\sigma}\;\de
h_{\mu\nu}} \right|_{h=0} + \frac{\de \chi_\la}{\de h_{\al\sigma}}
\;C^{\la\tau}\; \left. \frac{\de\chi_\tau}{\de
h_{\mu\nu}}  \right|_{h=0} \label{14}
\eeq
and $M_\al^\sigma$ is the operator of the $FP$ ghost action
\beq
M_\al^\sigma = \Box\,\delta^\sigma_\al+\na_\al
\na^\sigma
- 2 \,\be \; \na^\sigma \;\na_\al.
\label{15}
\eeq
The contributions of the operators $M_\al^\sigma$ and $C^{\mu\nu}$ to
the divergences of the effective action (\ref{13}) can be easily
evaluated by means of Barvinsky-Vilkovisky techniques \cite{bavi}.
The  result is
 \beq
- i \Tr \ln M_\mu^\nu = - \frac{1}{3\varepsilon}
\int d^4x \sqrt{-g}\;\left[
\left( - \frac45 + \frac1{8\be^2} + \frac1{2\be} \right)
R_{\tau\la}R^{\tau\la} +
\left( \frac35 + \frac1{16\be^2} \right)R^2  \right]
\label{16}
\eeq
and,
$$
- \frac{i}2 \Tr \ln C_\mu^\nu = - \frac{iN}2 \Tr \ln \left(\de_\mu^\nu\,
{\Box}\right) - \frac{i}2 \Tr \ln \left( g^{\mu\nu} \,\Box +
\ga \na^\mu \na^\nu - \na^\nu \na^\mu \right) \; =
$$
\beq
= - \frac{1}{3\varepsilon}
\int d^4x \sqrt{-g}\;\left[ \left( -\frac{4N}{5} +
\frac7{10}\right)\; R_{\tau\la}R^{\tau\la} +
\left(\frac{7N}{20} - \frac{3}{20} \right)\;R^2  \right].
\label{17}
\eeq
We remark that the last expression does not depend on $\ga$ due
to the cancellation first pointed out in Ref.
\cite{frts}.

The main technical problem is the calculation of
$\Tr \ln H_{\mu\nu,\al\sigma}$. $H$ is a differential operator of
$(2N+4)$ order, with coefficients depending on the
 curvature tensor of the background metric and its derivatives,  the
parameters of the gravitational action, and   the
gauge fixing parameters $\al,\be,\ga$. In spite of these difficulties
one can derive a general formula for the divergent part of $Tr \ln
H_{\mu\nu,\al\be}$ and then perform an explicit calculation for
the cosmological counterterm for arbitrary value $N$.
Let us introduce  dimensionful coupling constants
$$
\omega^i_n=m^{-2n}\;\alpha^i_{2n+4}
$$
by  absorbing the mass parameter in the dimensionless
couplings $\al^i_n$ of $S_{\rm eff}$. Since the
divergences do not depend on the explicit values of the gauge fixing
parameters, it is very convenient to fix their values to simplify
the calculations. A convenient choice is
\beq
\al = \frac{2}{\om^1_{N}},\;\;\;\;\;\;\;\;\;\;\;\;
\ga = -
\frac{2\om^2_{N}}{\om^1_{N}},\;\;\;\;\;\;\;\;\;\;\;\; \be =
\frac{\om^1_{N}}{4\om^2_{N}} + 1. \label{17i}
\eeq
For this choice of the gauge fixing, the operator becomes
\beq
&& H_{\mu\nu,\al\be} = \left( \frac{\om^1_N}{4} \;
\de_{\mu\nu ,}^{\;\;\;\;\;\rho\sigma} -
\frac{\om^1_N\;(\om^1_N+4\om^2_N)}{16\om^2_N}
\; g_{\mu\nu}\;g^{\rho\sigma}
\right)\; \left\{ \de_{\rho\sigma,\al\be}\,{\Box}^{N+2}  \right.
\cr
&&
\cr
&&+
{V_{\rho\sigma,\al\be}}^{\la_1 \la_2 \cdots \la_{2N+2}}
\na_{\la_1}\na_{\la_2}\cdots\na_{\la_{2N+2}} + \left.
{W_{\rho\sigma,\al\be}}^{\tau_1 \tau_2 \cdots \tau_{2N+1}}
\na_{\tau_1}\na_{\tau_2}\cdots\na_{\tau_{2N+1}}
\right.
\cr
&&\cr
&& \left.
+{U_{\rho\sigma,\al\be}}^{\up_1 \up_2 \cdots \up_{2N}}
\na_{\up_1}\na_{\up_2}\cdots\na_{\up_{2N}} + \CO(\na^{2N-1}) \right\}
\label{18}
\eeq
where $V, W, U$ depend on the dimensionful ratios like
${\om^1_{N - 1}}/{\om^1_N}$
and on the curvature tensor of the background
metric and its covariant derivatives.
Now we can use the  Barvinsky-Vilkovisky method
 \cite{bavi} to extract the divergences of $ \Tr \ln
H_{\mu\nu,\al\be}$. The  pre-factor of
(\ref{18}) does not contribute to the divergences and therefore it
can be omitted. Hence
$$
\Tr \ln H_{\mu\nu,\al\be}
= (N+2)\,\Tr \ln \left[ \de_{\mu\nu,\al\be}\, {\Box} \right]
+ \Tr {V_{\mu\nu,\al\be}}^{\la_1
\la_2 \cdots \la_{2N+2}}
\na_{\la_1}\na_{\la_2}\cdots\na_{\la_{2N+2}}\; \frac{1}{\Box^{N+2}}
$$
$$
+ \Tr {W_{\mu\nu,\al\be}}^{\tau_1 \tau_2 \cdots
\tau_{2N+1}} \na_{\tau_1}\na_{\tau_2}\cdots\na_{\tau_{2N+1}}\;
\frac{1}{\Box^{N+2}} + \Tr
{U_{\mu\nu,\al\be}}^{\up_1 \up_2 \cdots \up_{2N}}
\na_{\up_1}\na_{\up_2}\cdots\na_{\up_{2N}} \frac{1}{\Box^{N+2}}
$$
$$
- \frac12\; \de^{\rho\si ,\de\phi}\;
\Tr\, {V_{\mu\nu ,\rho\si}}^{\la_1  \cdots \la_{2N+2}}
\na_{\la_1}\cdots\na_{\la_{2N+2}} \frac{1}{\Box^{N+2}}
\;
{V_{\de\phi,\al\be}}^{\la'_1 \cdots \la'_{2N+2}}
\na_{\la'_1}\cdots\na_{\la'_{2N+2}} \frac{1}{\Box^{N+2}}
$$
\beq
+ \;(\rm terms\;\; of\;\; higher\;\; background\;\; dimension)
\label{19}
\eeq
Indeed by dimensional arguments
 the last higher  background  dimension terms do not contribute to
the divergences. The only  term which can not be  directly handled
by the method of \cite{bavi}, is that which involves
two  $V$ matrices. In particular there are new  terms
coming from the commutation of $V$ with the covariant derivatives
and the operators ${\Box}^{-1}$. However, since $V$ has dimensions
of curvature, and  derivatives of $V$ increases its
dimension, those terms do  not contribute
to the divergent part of the effective action. Therefore
as far as one loop divergences are concerned
  $V$ may be considered as a constant and we can
ignore its derivatives. The corresponding contribution can be written
as \footnote{ By the same dimensional reasons the
matrix $W$ of (\ref{18}) can not generate divergences and for such
a reason we shall not consider its contribution.}
\beq
 - \frac{1}{2}\; \de^{\up\psi,\de\phi}\Tr
{V_{\mu\nu\up\psi}}^{\la_1 \cdots \la_{2N+2}}
\;{V_{\de\phi\al\be}}^{\la'_1  \cdots \la'_{2N+2}}
\na_{\la_1}\cdots\na_{\la_{2N+2}}
\na_{\la'_1}\cdots\na_{\la'_{2N+2}} \frac{1}{\Box^{2N+4}}
\label{nu1}
\eeq
Now all the terms in the expansion admit the direct substitution of
the universal trace formulae of  Barvinsky-Vilkovisky and we can
derive  a general formula for the one-loop divergences of the theory.
To apply this formula  for a general $N$  one needs
to expand the action  up to second order in quantum fields, keeping
the first and second orders in  background curvature while
neglecting the higher orders and derivatives of the curvature.
The algebra involved in the calculation is quite cumbersome.
We will restrict ourselves to the calculation of the divergent
contributions to the cosmological constant counterterm. In order to
simplify the calculations we remark that  the background field
method gives the result which is valid for any background metric,
including the flat one. Thus the cosmological counterterm can be
derived for the flat background metric, which considerably
simplifies the calculation.  According to (\ref{13}) and (\ref{14})
the one-loop divergences are related with the Hessian  of the
classical action. If the background metric is flat, one can,
therefore, ignore all the terms which have more than two powers of
curvature, and  calculate the divergences of the theory
\beq
S =
\int d^4x \sqrt{-g}\;\left[ \sum_{n=0}^{N} \left( \om^1_{n}
\,R_{\mu\nu}\; \Box^n \;R^{\mu\nu} + \om^2_{n}\,
 R \;\Box^n \;R \right)
 -  \om_{-1} R +  \om_{-2}\right].
\label{20}
\eeq
The cosmological constant term $ \om_{-2}\sqrt{-g}$ does not give
divergent  contribution for any $N>0$, and Einstein term is relevant
for $N=1$ only. Let us first consider the case
of $N > 1$. The  relevant part of operator $H$ for the divergent
contribution has the form
\beq
H_{\mu\nu,\al\be} = \de^{\mu\nu,\al\be}\;\Box^{N+2}+
{V_{\mu\nu,\al\be}}^{\la\tau\rho\sigma}
\Box^{N-1} \na_\la\na_\tau\na_\rho\na_\sigma
+ {U_{\mu\nu,\al\be}}^{\la\tau\rho\sigma}
\Box^{N-2} \na_\la\na_\tau\na_\rho\na_\sigma
\label{op}
\eeq
Using well known expansions of $R_{\mu\nu}$
we obtain the following expressions for $V$
\beq
&&\left[V_{\mu\nu,\al\be}
\right]^{\la\tau\rho\sigma}
= \frac{\om^1_{N-1}}{\om^1_N} \;\de_{\mu\nu,\al\be}g^{\la\tau}g^{\rho\sigma}
+\frac{\om^1_N \om^2_{N-1} - \om^2_N \om^1_{N-1}}{\om^1_N (\om^1_N +
3\om^2_N)}\;
g_{\mu\nu} g_{\al\be} g^{\la\tau}g^{\rho\sigma}\cr
&&
+
\frac{2(\om^1_{N-1}+2\om^2_{N-1})}{\om^1_N}\;
\de_{\mu\nu,\al\be}\de^{\la\tau,\rho\sigma}
+\frac{\om^2_N\om^1_{N-1} - \om^1_N\om^2_{N-1}}{\om^1_N (\om^1_N + 3\om^2_N)}\;
g_{\mu\nu}{\de_{\al\be}}^{\la\tau}g^{\rho\sigma}\cr
&&
-\frac{\om^1_{N-1}}{2\om^1_N}\;g^{\la\tau}\;
\left(
g_{\mu\al}\de_{\be}^{\sigma}\de_{\nu}^{\rho}+
g_{\nu\al}\de_{\be}^{\sigma}\de_{\mu}^{\rho}+
g_{\mu\be}\de_{\al}^{\sigma}\de_{\nu}^{\rho}+
g_{\nu\be}\de_{\al}^{\sigma}\de_{\mu}^{\rho}
\right)
\cr
&&
- \frac{\om^1_{N-1}+4\om^2_{N-1}}{\om^1_N}\;
g_{\la\tau}g_{\al\be}{\de_{\nu\mu}}^{\rho\sigma} \label{21a}
\eeq
\vskip 0.5mm
and for $U$
\vskip 0.5mm
\beq
&&
\left[U_{\mu\nu,\al\be}
\right]^{\la\tau\rho\sigma}
= \frac{\om^1_{N-2}}{\om^1_N} \;\de_{\mu\nu,\al\be}g^{\la\tau}g^{\rho\sigma}
+\frac{\om^1_N\om^2_{N-2} - \om^2_N\om^1_{N-2}}{\om^1_N (\om^1_N + 3\om^2_N)}\;
g_{\mu\nu}g_{\al\be}g^{\la\tau}g^{\rho\sigma}\cr
&&
+\frac{2(\om^1_{N-2}+2\om^2_{N-2})}{\om^1_N}\;
\de_{\mu\nu,\al\be}\de^{\la\tau,\rho\sigma}
+\frac{\om^2_N\om^1_{N-2} - \om^1_N\om^2_{N-2}}{\om^1_N (\om^1_N + 3\om^2_N)}\;
g_{\mu\nu}{\de_{\al\be ,}}^{\la\tau}g^{\rho\sigma}\cr
&&
-\frac{\om^1_{N-2}}{2\om^1_N}
g^{\la\tau}
\left(
g_{\mu\al}\de_{\be}^{\sigma}\de_{\nu}^{\rho}+
g_{\nu\al}\de_{\be}^{\sigma}\de_{\mu}^{\rho}+
g_{\mu\be}\de_{\al}^{\sigma}\de_{\nu}^{\rho}+
g_{\nu\be}\de_{\al}^{\sigma}\de_{\mu}^{\rho}
\right)
\cr
&&
- \frac{\om^1_{N-2}+4\om^2_{N-2}}{\om^1_N}
g_{\la\tau}g_{\al\be}{\de_{\nu\mu ,}}^{\rho\sigma} \label{21}
\eeq
Using the \BV\ trace formulae in (\ref{19}), for
  flat background metric, we get the following expression for the
divergences
\beq
\frac{i}{2} \Tr \ln H_{\mu\nu,\al\be}  =
\frac1{\varepsilon} \int d^4 x \sqrt{-g} \;\tr\; \left(
-\frac1{12}\; U^{\la\tau\rho\sigma}\,g^{(2)}_{\la\tau\rho\sigma} +
\frac1{1920}\; V^{\la\tau\rho\sigma} V^{\la'\tau'\rho'\sigma'}\,
g^{(4)}_{\la\tau\rho\sigma\la'\tau'\rho'\sigma'} \right),
\label{22}
\eeq
where $\varepsilon = (4 \pi)^2 \,(n-4)$ is the parameter of the dimensional
regularization, and \cite{bavi}
$$
g^{(2)}_{\la\tau\rho\sigma}=g_{\la\tau}\,g_{\sigma}+
g_{\la\rho}\,g_{\tau\sigma}+g_{\la\sigma}\,g_{\rho\tau}
$$
$$
g^{(4)}_{\la\tau\la '\tau '\rho\sigma\rho '\sigma '} =
g_{\la\tau}\,g_{\la '\tau '}\,g_{\rho\sigma}\,g_{\rho '\sigma '} +
{\rm permutations}.
$$
 Now, taking into account
(\ref{21a})--(\ref{22}) we calculate the one-loop divergences for
the theory  on flat background. The result is
\beq
\Gamma^{div}_{N} =
\frac1{\varepsilon} \int d^4 x \sqrt{-g} \; \left[
-\frac2{\om^1_N(\om^1_N + 3\om^2_N)}\; u(N)
 + \frac1{(\om^1_N)^2(\om^1_N + 3\om^2_N)^2}\; v(N) \right]
\label{23}
\eeq
where
\beq
u(N,N>1)&=& (6\om^1_N\;\om^1_{N-2} +
15 \om^2_N\;\om^1_{N-2} + 3\om^2_{N-2}\;\om^1_N)
\cr &&
\cr
v(N) &=& (\om^1_N)^2 (\om^1_{N-1} + 3\om^2_{N-1})^2 +
5(\om^1_{N-1})^2(\om^1_N + 3\om^2_N)^2 .
\label{24}
\eeq
For the case $N=1$ one needs to change the
expression for $U$ in (\ref{21}).
\beq
{U_{\mu\nu,\al\be}}^{\la\tau\rho\sigma}(N=1)
&=& g^{\la\tau} \,\om_{-1}\,\left[\,
-\frac{1}{\om^1_1 } \;\de_{\mu\nu,\al\be}\,g^{\rho\sigma}
+\frac{\om^1_1 + 2\om^2_1}{2\om^1_1 (\om^1_1 + 3\om^2_1) }\;
g_{\mu\nu}\,g_{\al\be}\,g^{\rho\sigma}
\right.
\cr  &-&
\left.
- \frac{(\om^1_1+2\om^2_1)}{2\om^1_1
(\om^1_1 + 3\om^2_1) }\;
g_{\mu\nu} \,\de_{\al\be}^{\,\;\;\;\rho\sigma}
- \frac{1}{\om^1_1 }\;
g_{\al\be}\,\de_{\nu\mu}^{\,\,\,\,\rho\sigma}
\right.
\cr
&+& \left.\frac{1}{2 \om^1_1 }\;
\left(
g_{\mu\al}\de_{\be}^{\sigma}\de_{\nu}^{\rho}+
g_{\nu\al}\de_{\be}^{\sigma}\de_{\mu}^{\rho}+
g_{\mu\be}\de_{\al}^{\sigma}\de_{\nu}^{\rho}+
g_{\nu\be}\de_{\al}^{\sigma}\de_{\mu}^{\rho}
\right)\,\right]
\label{25}
\eeq
Then,
after inserting this expression into the formula (\ref{22}) we
also obtain (\ref{23}) but with a different coefficient $u(N=1)$.
\beq
u(N=1) =  \frac{3}{2}\;\om_{-1}\;\left(3\,\om^1_1 + 10\,\om^2_1 \right)
\label{26}
\eeq
This formulae (\ref{23}) (\ref{24}) and (\ref{25})
complete the calculation of the cosmological
constant counterterm
in an effective gravity theory $S_{\rm eff}$. For $N>2$ the above expressions
are
exact since there are no any additional divergences for the
cosmological constant at higher loops.
It is important to recall that the  result is
independent on the choice of the gauge fixing condition and on the
parametrization of the quantum field. In this respect, it differs
from the counterterms which appear in quantum gravity with
two derivatives \cite{ktt}
and four derivatives \cite{frts,avr}. Thus
adding the higher derivative terms to the effective theory
introduces some relevant difference from this viewpoint.

It should be also interesting to calculate
the divergence of the coefficient of the Einstein term $R$,
because the corresponding beta function $\be_{G}=\be\om_{-1}$
describes the running of the gravitational constant with the change
of energy scale. The formulas (\ref{19}) and (\ref{nu1}) enable one to perform
such a calculation for any finite $N$, but it would require an effort beyond
our present possibilities.
By dimensional arguments the counterterm which is linear in curvature
has to be a linear combination of the coefficients
$ \,\om^1_{N-1}, \om^2_{N-1}$.
We know that all the coefficients entering in  this
linear combination will give a factor
${\om^1_N}^{-1} \;(\om^1_N + 3\om^2_N)^{-1}$, but the exact value requires
explicit calculation.
The renormalization of the parameters $\om^1_0, \om^2_0$ of the four
derivative terms (\ref{9}) can be also obtained from (\ref{19}) and
(\ref{nu1}). They are bilinear in $(\om^1_{N}, \om^2_{N}, \cdots)$ and
contain the factor $({\om^1_N})^{-1} \;(\om^1_N + 3\om^2_N)^{-1}$.

The logarithmic divergent contribution to the renormalization
of the cosmological constant (\ref{23}) yields a
renormalization group equation which governs its dependence on
the energy scale. The corresponding beta-function is
\beq
\be_{\om_{-2}} = \frac{1}{(4\pi)^2}\;
\left[ - \frac{2\;u(N)}{\om^1_N(\om^1_N + 3\om^2_N)} +
\frac{v(N)}{(\om^1_N)^2(\om^1_N + 3\om^2_N)^2}
\right].                                       \label{bella}
\eeq

One interesting consequence of this general analysis is that it provides
different scaling scenarios for the running of the cosmological constant
(as well as for the other relevant couplings). In principle it is possible
to impose more constraints on those coefficients to suppress all
logarithmically
divergent contributions. In such a case the corresponding theory is finite
and all beta functions vanish. But this behaviour only holds for energies
higher than the effective scale of the theory. When the energy moves
to lower values the leading ultraviolet terms
with $2N+4$ derivatives become irrelevant and    new scaling
scenarios  (\ref{bella}) governed by the terms with $2N+2$ derivatives
emerge.
 This can be seen from
our calculations taking the limit when the leading ultraviolet
coefficients tend to zero. We recover in such a case the beta
function corresponding to the subleading terms. In this way
the theory provides a hierarchy of different scaling scenarios from
very short distances to cosmological distances governed by different
beta functions. However, properly speaking to implement such multiple
scale scenario we will need a set of different dimensionful
parameter scales $\mu_1, \cdots ,\mu_N$.
One interesting feature of this scenario is the existence of
 crossover between the different
scaling windows appears. The renormalization group flow runs from
ultraviolet  fixed points to infrared fixed points step by step in a
continuous non-linear way.

 In this sense an open approach to quantum gravity
it is possible, although the compatibility with closed approaches
coming from some fundamental theories is also possible by imposing
the corresponding constraints for the different parameter of the
effective action $S_{\rm eff}$. In particular, if a non-trivial fixed
point is found the possibility of having a non-perturbative
approach to quantum gravity is open. The theory defined by scaling
limit around this point could have unexpected non-perturbative
effects.

\vskip 6mm

\noindent
{\large \bf 4. Structure of mass poles}
\vskip 2mm

In the previous section we have seen that the  high
derivative theory  provides an interesting smooth approach to
quantum gravity. In this framework the effective theory can exhibit
interesting features like the vanishing of the coupling constant in
the ultraviolet regime. Moreover, the theory provides a series of
different scaling scenarios for all relevant physical quantities
governed by different renormalization group fixed points.

However, as it is well known  such a versatility of
high derivatives theories
implies  serious problems from an unitarity viewpoint.  In
particular, the theory contains
a plethora of unphysical
massive ghosts. The
analysis of this problem in the framework of string
effective models \cite{zwe,dere} have shown that the massive ghosts
disappear from the spectrum in a special parametrization of the
metric $g_{\mu\nu}$ (the background metric in its
target space), while from string theory point of view this
parametrization does not differ from the others \cite{tse}.

Recently,  it was suggested  that in   theories with
more than two  derivatives of the metric it could be possible
to find a ghost scenario with only  one ghost which would be the
most massive fundamental particle of the theory having negative
norm states in the Hilbert space, whereas the other
lighter fundamental particles would have positive norm states
which are compatible with unitarity \cite{sha}.
In fact, in the theories of the type considered in
the present paper the
 structure of the (euclidean) propagator in the spin-2 and spin-0
reads
\beq G(k) = \left( l_{2N+4} \, k^{2N+4} + l_{2N+2}\,
k^{2N+2} + l_{2N}\, k^{2N} + \cdots + l_{2} \, k^{2} \right)^{-1}
\label{27}
\eeq
where $l_{i}$ are real
numbers related with the values of $\om^1_i,\; \om^2_i$ coefficients
in (\ref{20}). The expression (\ref{27}) can be
decomposed in terms of simple propagators as
\beq
G(k) = \frac{A_{0}}{k^2} +
 \frac{A_{1}}{k^2 + m_1^2} + \frac{A_{2}}{k^2 + m_2^2}
+ \cdots + \frac{A_{N+1}}{k^2 + m_{N+1}^2}
\label{28}
\eeq
where  the masses $m_j^2$ can be real or complex depending on the
values of the coefficients $l_i$ (\ref{27}). The complex masses
are always grouped in conjugate pairs. The idea introduced
in Ref. \cite{sha}
was based on the assumption that
 for some physical values of $l_i$ all
the mass poles $m_j^2$ are real and positive, $ 0 < m_1^2 < m_2^2 <
m_3^2 < \cdots < m_{N+1}^2$, and the coefficients $A_r$
are all positive $A_r > 0$ for $r = 0,1,\cdots,N$, except the last
one $A_{N+1}$ which is negative $A_{N+1} < 0$. In this case only the
heaviest particle would be  an
unphysical ghost, whereas all the  others would be
 ordinary massive
particles with positive energy.
Thus,  the
effective theory would be  unitary only till    energies of
the order of the mass  of the heaviest unphysical particle
$m^2_{N+1}$, although this mass can be chosen much larger than
the masses of other particles.   Beyond such a scale the theory
becomes unphysical.  Unitarity can  only be completely restored
for all energy scales in the ultimate fundamental theory.

Unfortunately, this scheme can not
work, although from a theoretical viewpoint was very appealing. The
problem is that the assumptions concerning the behaviour of the
massive poles $m^2_i$, and their residua $ A_j$ in the high
derivative theory  are not consistent. The reason  is that
(see Appendix A for details)
for any real monotone sequence of masses
 $0 < m_1^2 < m_2^2 < m_3^2 < \cdots < m_{N+1}^2$, the signs of the
corresponding
residua  alternate, i.e.
$\sign\ [A_j] = -\; \sign\ [A_{j+1}]$. This can be easily
understood from the
the basic property of any polynomial
with real coefficients and real zeros which establishes
that the signs of the slopes at the
zeros do always alternate between  consecutive zeros.
Hence the physical assumption made in Ref. \cite{sha} is
never satisfied
for real masses. In the case of complex poles we have an
even more pathological situation. Complex masses $m^2<0$
lead to pairs of unphysical tachyons  and
for the remaining   real masses the above argument
show  that they correspond  to  alternating pairs of
particles and ghosts. However, this pathological ghost masses
distribution does not exclude the existence
of a spectrum changing field reparametrization  mapping the theory
into an unitary theory \cite{aa2}.

In any case the dynamical role of unphysical massive ghosts only appears
 at energies of  Planck order \cite{ste}.
At   very low energies (corresponding to the macroscopic length)
they do not propagate, and the only  relevant excitation
is that related with the massless  graviton. In particular,
all infrared  quantum effects of the high
derivative theory   are the same as in Einstein gravity.
This is specially relevant for quantities with singular infrared
behaviour, e.g. long range correlation functions.
For instance, if we evaluate the one loop
quantum corrections to the gravitational potential, we get the
same result as Donoghue
 \cite{don}, independently of the details of the higher
derivative theory  we consider. This remarkable result is due to
the universality of the non-local singular infrared
contributions to correlation
functions \cite{jl}. Now, in the pure  Einstein theory
due to the absence of a mass term this singular
IR behaviour extends to the ultraviolet and in fact can be
easily computed in the UV regime, where it is linked with the
divergent behaviour of the counterterms.
The relation is similar to the one which exists  between the UV
counterterms and the beta functions of the coupling constant. Now,
the UV behaviour of the higher derivative gravity is completely
different and it is obviously dependent on the details (coefficients
and number of derivatives) of the effective action. The
renormalization group has in such a case a non-linear behaviour which
goes from the particular UV behaviour associated to our regularized
theory to the universal infrared Einstein regime. Between these two
regimes we have momenta domains where new  regimes with different
scaling  behaviour appear provided the  masses of the different
massive ghosts  of the theory are widely separated. The nonlinear
behaviour of the renormalization group allows to go from the UV to
the infrared through this series of unstable regimes. The relevant
result is that the long distance behaviour of the theory is the same
as Einstein gravity with some small quantum corrections
\cite{don}.  Indeed the contributions of the higher derivative terms
become more and more  relevant as we increase the range of energies
or what it is the same we go to shorter distances.
One of their effects is the generation of a running of the gravitational
constant with the change of energy scale, and a running of the
cosmological constant which can become stable at short distances
due to the vanishing of the corresponding counterterms.

\vskip 6mm

\noindent
{\large \bf 5. Interaction with matter fields}
\vskip 2mm

Let us now explore the interaction of high derivative gravity
with  matter and gauge fields. In particular, it
is interesting to analyze how the quantum effects of this theory can
change the scaling properties of the matter
field correlation functions and  the
effective potential of the Higgs fields. For that purpose we
introduce  a general gauge model containing spinor, vector and
scalar Higgs fields with gauge, Yukawa and scalar interactions. The
total action has the form
\beq
S_{\rm total} = S_{\rm eff} + S_{\rm  gauge}
\label{30}
\eeq
where $S_{\rm eff}$ denotes the action of our high derivative gravity
(\ref{1})  and $S_{\rm gauge}$
is the action of the gauge model. We assume, as in the previous
sections, that the gravitational propagator has an ultraviolet
behaviour of order $\CO \left({k^{-(2N+4)}} \right)$.

The diagrams contributing to any correlation function  with external
matter lines split into two sets. Diagrams with only matter
internal lines, and  diagrams with an internal
gravitational propagators. The contributions
of the diagrams of the first type
(e.g.  diagrams of the Figure 1) are equal to the contributions
of the gauge model in a classical gravitational background.
It can be shown \cite{book} (and references therein)
that in general the corresponding beta-functions for the
matter fields couplings (gauge, Yukawa and scalar) are exactly
the same as in flat space-time.
The novel aspects are the need of the   non-minimal coupling
$\xi R \phi^2$ for every scalar field $\phi$, the
renormalization of $\xi$ and
the appearance of the corresponding beta-function $\be_\xi$ \cite{book}.

Indeed, there also appear  additional contributions to the
renormalization of the parameters
$\om^1_0,\;\om^2_0,\om_{-1},\;\om_{-2}$ generated by the loops of
matter fields in the diagrams with only gravitational external lines
(e.g. diagram (2) of Figure 1).
Diagrams involving only  internal and external gravitional lines
are only divergent when the inequality  (\ref{8}) holds. Both types of
divergent counter-terms are of the form (\ref{9}).

 Let us now consider the most general
 diagrams which have  some external
matter lines and any kind of internal
lines (Figure 2).
Here one has to consider the cases $N>1$ and $N=1$ separately.
 For $N>1$ all such diagrams are finite by power counting,
thus they do not contribute to the beta-functions for the matter
field couplings (gauge, Yukawa, scalar), which are actually the same
as in the theory without quantum gravity.
For gauge coupling constants the result is even more general. Their
 beta-function are not affected by any kind of the gravitational
  interaction. In the Einstein gravity it was
shown in Ref. \cite{dene}. The effect is essentially based on the
non-renormalizability of the Einstein-Maxwell or Einstein-Yang-Mills
system (power counting does not
permit any gravitational contribution  to ${(F^a_{\mu\nu}})^2$). In
the fourth derivative quantum gravity the gravity-independence of
the gauge coupling has been found in Ref. \cite{frts} as a result of
unexpected cancellation of two diagrams. What we have shown  is that
the same result holds for the general theory (\ref{1}) with $N>0$.
In summary, asymptotic freedom is stable
under quantization of gravity.

In the case $N=1$, there appear two new divergent
diagrams with external scalar lines and internal graviton loops
 (see Figure 3). The divergence   of these diagrams is generated by
the nonminimal interaction term in the scalar (Higgs) sector of the
gauge model. In fact the divergences generated by such mixed diagrams
are always universal, i.e.   they do not depend on
the gauge group or multiplet composition. Consider, for
simplicity, one real scalar field. \beq
S_{sc} = \int d^4x\sqrt{-g}\,\left(
\frac12\,g^{\mu\nu}\,
\partial_\mu \phi\,\partial_\nu\phi + \frac12\,m^2\,\phi^2
+ \frac12\,\xi\,R\,\phi^2 - \frac{f}{24}\,\phi^4 \right)   \label{mat1}
\eeq
Diagrams of Figure 3 can
generate UV divergences  only  if the two derivatives introduced by
the scalar curvature in $\frac12\,\xi\,R\,\phi^2$ term act to the
internal gravitational lines. Therefore they can   only affect
the  renormalization of the mass $m^2$ but none of the
coupling constants.

The corresponding mass counterterm is (see Appendix C)
$$
\Ga^{\rm div}_{\rm scalar} = \frac{1}{\varepsilon}\,
\frac{3\,\xi}{\om_1^1+3\,\om_1^2}
\,\left( \frac{3\,\om_1^1 + 10\,\om_1^2}{\om_1^1} - \xi \right)\int
d^4x\sqrt{-g} \; \phi^2\; .
$$
This divergence generates a gravitational correction to the known
beta-function for the scalar mass, which appears due to the gauge,
Yukawa and scalar interactions
$$
\beta^{\rm total}_{m^2} = \beta^{\rm known}_{m^2} +
\de^{\rm grav}\,\beta_{m^2} $$
\beq
\de^{\rm grav}\,\beta_{m^2} =
\frac{1}{16\,\pi^2}\,\frac{3}{\left(\om^1_1 + 3\om^2_2\right)}\,\xi\,
\left( \xi - \frac{3\,\om^1_1 + 10\,\om^2_1}{\om^1_1}\right)
\label{mat2}
\eeq

In summary, in the framework of the higher derivative model of quantum
gravity the beta functions for the gauge, Yukawa and scalar
self-coupling are not affected by quantum gravity
corrections. The only   quantum
effect is the correction to the mass beta-function of the scalar field
(\ref{mat2}) which only appears  for the case $N=1$.

 We remark that in
fourth derivative quantum gravity the contributions of gravitational
loops to the beta-functions for the Yukawa and scalar couplings are
nontrivial \cite{bush}, they can even change the asymptotic
behaviour of such theories (see also \cite{book}). In contrast, for
$N>0$ there is no divergent contribution of gravity to the matter
or gauge fields sector.

\vskip 6mm

\noindent
{\large \bf 6. Conclusions}
\vskip 2mm

The above results show that in the high derivative approach to
quantum gravity the UV behaviour is so smooth that its corrections
to gauge theories do not modify
the scaling UV behaviour of gauge and matter correlation functions
for $N>0$.
However, in general, the pure gravitational selfinteraction is not
 completely self-regularized. As we have shown one-loop divergences
are generically
of the same type (\ref{9}) that those of the Einstein theory. This is a
general feature of non-abelian gauge theories. The analysis of this
one-loop divergences shows that they can be cancelled by any of the
following prescriptions:

\begin{itemize}
\item{i)} Pauli-Villars covariant regularization

\item{ii)} Fine tuning of the high derivative couplings

\end{itemize}

The theory with a finite number of couplings can be considered
either as a regularization of quantum gravity or better as an
effective theory of quantum gravity. As effective theory some of
its infrared properties, like the corrections to the Newtonian
potential, are universal and do not depend on the UV behaviour of
the theory. We have shown that the structure of mass poles is such
that the ghost masses are intercalated between those of particles
in such a way that does not allow to recover unitarity unless we
restrict to energies below the first mass scale of the theory.

Another interesting feature of the high derivative approach
is that the UV scaling regime is independent of the gauge
parameter and the parametrization of metric fields.

In the cases where the theory is finite for all momentum scales
it might be possible, in principle, to find interesting
non-perturbative scaling  regimes which shall give rise to novel
non-perturbative approaches to quantum gravity.

In summary, the high derivative approach offers a good
selfconsistent framework to deal with quantum gravity effects
for intermediate scales between low energy phenomenology and high
energy fundamental unified theories.

 \vskip 6mm
\noindent{\large \sl Acknowledgments}

We acknowledge  S. Ketov for helpful comments on
the reparametrization ambiguity in string theory.
I.Sh. is grateful to  MEC-DGICYT for a fellowship and
Departamento de Fisica--ICE,
Universidade Federal de Juiz de Fora, MG--Brazil,
for kind hospitality. This work is partially supported
by  CICYT (grant AEN94-0218) and INTAS q(grant 93-1038) .

\vskip 6mm
{\large \bf Appendix A}
\vskip 2mm

Let us prove now the intercalation  property of
particles an ghost masses which have been used in the main text. We
consider for simplicity the euclidean formalism. The euclidean
propagator $G(k^2)$  has the form
$$
G(k^2)={l_{2N+4}^{-1}\over \prod_{i=0}^{N+1}(k^2+m_i^2)} \eqno{(A.1)}
$$
where the masses $m_j$, $j=0\cdots N+1$ are functions of the
coefficients $l_{2j+2}$;
and can be rewritten in the form (\ref{28}). Therefore,
$G(k)$ has simple poles in $k^2=-m_j^2$, $j=0\cdots N+1$ if all the masses are
different.

In the complex plane ${\IC}$,
$G(z)$ is a meromorphic function with $N+2$ simple
poles. Let $\Gamma_j$ be a closed path in ${\IC}$
around the pole $-m_j^2$ not encircling any other pole.
Then, from (A.1), we have, for $j=0\cdots N+1$  %
$$
\oint_{\Gamma_j}G(z)=2\pi i{\rm Res}\left
[{l_{2N+4}^{-1}\over \prod_{i=0}^{N+1}(z+m_i^2)}, -m_j^2\right]=
2\pi i l_{2N+4}^{-1}
\prod_{i=0,i\ne j}^{N+1}{1\over (m_i^2-m_j^2)} \eqno(A.2)
$$
and from (\ref{28})
$$
\oint_{\Gamma_j}G(z)=2\pi i{\rm Res}\biggl
[\sum_{i=0}^{N+1}{A_i\over (z+m_i^2)}, -m_j^2\biggr]=
2\pi iA_j,\hskip 1cm j=0\cdots N+1 \eqno(A.3)
$$
Then, from (A.2) and (A.3) we get%
$$
A_j=l_{2N+4}^{-1}\prod_{i=0,i\ne j}^{N+1}{1\over (m_i^2-m_j^2)},
\hskip 1cm j=0\cdots N+1 \eqno(A.4)
$$
When the masses $m_j$ are real or purely imaginary,
their squares  $m_j^2$ are real, and once they are ordered
($m_i^2<m_{i+1}^2 \hskip 1mm\forall\hskip 1mm i=0\cdots N$),
(A.4) tells us that the signs of the residua  alternate,
i.e. sign$\lbrack A_j\rbrack =(-1)^{j} $sign$\lbrack
l_{2N+4}^{-1}\rbrack$, $j=0\cdots N+1$.

This result can also be understood in pure algebraic terms. If the
polynomial $G(k^2)^{-1}$ of $k^2$ which defines the propagator only
has non-degenerate real zeros (and this is the assumption we have
made)  the signs of the slopes at two consecutive zeros alternates.

On the other hand, if one $m_j^2$ is complex,  there is another
one $m_{l}^2$ which is its complex conjugate,
$m_{l}^2={m_j^2}^{*}$.
We can easily show that their residua $A_j$, $A_{l}$
are also pairs
of complex conjugate numbers  $A_j={A_{l}}^{*}$, whereas
the residua $A_j$ corresponding to the remaining  real masses
$m_j^2$ have alternating signs.

\vskip 6mm
{\large \bf Appendix B}
\vskip 2mm

Let us prove in detail the gauge fixing independence of the
one-loop divergences by using
the method of \cite{bavi} (see also  \cite{avr,shja}
for the fourth derivative gravity).

Diffeomorphism invariance implies that the classical action is
invariant under the infinitesimal transformation
$$
\delta g_{\mu\nu} = {\cal D}_{\mu\nu,\la}\xi^{\la},
\,\,\,\,\,\,\,\,\,\,\,\,\,\,\,\,
{\cal D}_{\mu\nu,\la} = - \left( g_{\mu\la}\na_\nu + g_{\nu\la}\na_\mu \right)
$$
which leads to the Noether identity
$$
{\cal D}_{\mu\nu,\la}\,{{\delta S}\over{\delta g_{\mu\nu}}}
 = 0
\eqno(B.1)
$$
The gauge fixing term (\ref{6i}) introduces a number of gauge
parameters $\sigma_i,\gamma_i,\beta$. Let us generically denote
them by $\sigma$.
The propagators $G^{\mu\nu,\al\be}$, ${\Om^{\al}}_{\be}$ of the metric
field $h_{\mu\nu}$ and of
the Faddeev-Popov ghosts $\bar{c}_{\be_1} ,c^{\be_2}$  are defined
as usual by the relations
$$
K_{\mu\nu, \al\be}\;G^{\al\be,\rho\sigma} =
{{\delta}_{\mu\nu}}^{\rho\sigma},\;\;\;\;\;\;\;\;\;\;\;\;\;\;\;\;
{M^{\al}}_{\beta}\;{\Om^{\beta}}_{\lambda} =
{\delta}^{\al}_{\lambda}                 \eqno(B.2)
$$
where
$$
K_{\mu\nu,\al\be} = {{\delta}^2 S \over {{\delta} g_{\mu\nu}\,
{\delta}g_{\al\be}}} + { {\delta\chi_{\rho}} \over {\delta g_{\mu\nu}} }\;
C^{\rho\sigma}\; { {\delta\chi_{\si}} \over {\delta g_{\al\be}} }
$$
and
$$
{M^{\al}}_{\be} =  C^{\al\ga}\;
{{\delta\chi_{\ga}} \over {\delta g_{\mu\nu} }}\;
{\cal D}_{\mu\nu,\be}                             \eqno(B.3)
$$

{}From (B.1) and (B.3) it follows the Ward identity which links
the two
propagators and the gauge fixing  functional
$$
{\cal D}_{\mu\nu,\be} \; {\Om^{\be}}_{\al}
- G_{\mu\nu,\rho\si}\;
{ {\delta\chi_{\al}} \over {\delta g_{\rho\si} } } =
- G_{\mu\nu,\rho\si}\; {\delta {\cal D}_{\la\tau,\ga} \over {\delta
g_{\rho\si}}}
{{\de S} \over { \de g_{\la\tau} }}\; \;
{\Om^{\ga}}_{\al}                                  \eqno(B.4)
$$
The one loop effective action $\Gamma$ is given by the
expression
$$
i \Gamma = - {1\over2}\,\Tr\ln\;K_{\mu\nu,\al\be} +
\Tr\ln\;{M_{\al}}^{\beta} - {1\over2}
\Tr\ln\;C^{\al\be}                      \eqno(B.5)
$$
Notice that we use a different definition of $M$ and that is why the
sign of $ \Tr\ln\; C$ in  (B.5) is opposite to that of expression (\ref{13}).
If we denote by $\dot{X}$ the derivative of the quantity $X$
with respect to a generic parameter introduced by the gauge fixing
functional, then the derivative of (B.5) with respect of such a
parameter can be estimated with the use of (B.4). The result is
 $$
i\dot{\Gamma}= -{1\over 2}\;\Tr\,
G_{\mu\nu,\rho\si}\;
{\delta {\cal D}_{\la\tau,\be} \over{\delta g_{\rho\si} }}\;
{\delta S \over \delta g_{\la\tau}}\;
{\Om^{\be}}^{\al}\;
\left[
{ {\delta \chi_{\ga}} \over {\delta g_{\mu\nu}}}\;
{\dot{C}}^{\al\ga}  + 2\, C_{\al\ga}\;
\dot{ { \delta \chi^{\ga}} \over {\delta g_{\mu\nu}} }
\right ]                                              \eqno(B.6)
$$

The difference between the one-loop corrections in two different
gauges,
${\Gamma}({\sigma}) - {\Gamma}({\bar \sigma})$
is obtained by integrating
(B.6)  for all the gauge fixing parameters from $\sigma$
to $\bar{\sigma}$
 \cite{shja}.
The equation (B.6) gives the general form of the gauge
parameters dependence of the one-loop effective action which is proportional
to the motion equations, ${\delta S} / \delta g_{\al\be}.$
For the higher derivative theory   with   $N>0$ this
implies the independence of the one loop divergent contributions
on the gauge parameters ${\sigma_n}, \gamma_n$ and $\beta$
as proved in section 3.

\vskip 6mm
{\large \bf Appendix C}
\vskip 2mm
The   divergent part of the diagrams of Figure 3
can be calculated by means of  the background field method and
Schwinger-DeWitt technique, using the generalization introduced in Ref.
\cite{bavi} and  the methods developed in  Ref. \cite{bush} (see
also \cite{book}). In the present theory
the calculation is technically  simpler than
the one carried out in \cite{bush}.
First, we notice that, by power counting, we only need
 to calculate the divergent terms of the form $Z\,\phi^2$ where
$\phi$ is the background scalar, and $Z$ is a divergent coefficient
with dimension of ${\rm mass}^2$. Besides the splitting of
the metric $g_{\mu\nu} \rightarrow g_{\mu\nu} + h_{\mu\nu}$ into
background and quantum fluctuations we have a similar  splitting for the
scalar field $\phi \rightarrow \phi + \varphi$.
 We can therefore consider the
background field $\phi$ as a constant.
The second diagram contains purely gravitational loops,
and   it can be evaluated with the algorithm used
in (\ref{23}). Then one gets the result by replacing $$
\om_{-1} \longrightarrow \om_{-1} - \frac12\,\xi\,\phi^2   \eqno(C.1)
$$
in the
expression (\ref{26}).

The calculation of the divergences generated by the first diagram is more
involved. Power counting tells us that this contribution
 only appears   when all derivatives of the vertices of those digrams
 act on the internal lines. In the framework of the background field method
this means that we need to keep only the terms with two derivatives of the
quantum fields when performing the background field
expansion of the $\xi\,R\,\phi^2$ term in (\ref{mat1}).
On the other hand one can neglect  all the non-leading order terms
in  the pure gravitational $h-h$ sector and
consider only the contribution of the higher derivative terms,
because all the others do not contribute to the divergent
counterterms. The same feature occurs for non-leading derivative
terms in the scalar $\phi-\phi$ sector. However the higher
order terms in the mixed $\phi-h$ and $h-\phi$ sectors are relevant.
In summary, it can be shown that all the divergences are contained in
the expression
 $ (i/2)\,\Tr\,\ln\,{\hat H} $
where
$$
{\hat H} =
\left(\matrix{
A_{\mu\nu,\al\be}\,{\Box}^3
&  \left[ P^{\la\tau} \right]_{\mu\nu}\,\na_\la\,\na_\tau
\cr
\left[ Q^{\la\tau} \right]_{\al\be} \,\na_\la\,\na_\tau
&  - \frac12\;{\Box}                 \cr}
    \right)         ,                     \eqno(C.2)
$$

and $A,P,Q$ are given by
$$
  A_{\mu\nu,\al\be}
= \frac{\om_1^1}{4}\;
\left(\de_{\mu\nu,\al\be} - \frac{\om_1^1+4\om_1^2}{\om_1^2}\,
 g_{\mu\nu}\,g_{\al\be} \right)
$$
$$
\left[ P^{\la\tau}\right]_{\mu\nu}
= \frac12\,\xi\,\phi\,
\left({\de^{\la\tau}}_{\mu\nu} - g^{\la\tau}\,g_{\mu\nu} \right)
,\,\,\,\,\,\,\,\,\,\,\,\,\,
\left[{\hat Q}^{\la\tau}\right]_{\al\be}  = \frac12\,\xi\,\phi\,
\left({\de^{\la\tau}}_{\al\be} - g^{\la\tau}\,g_{\al\be} \right)
\eqno(C.3)
$$
This formula can be compared with the similar
expression associated to  fourth derivative gravity coupled to
matter \cite{book}.

Thus,
$$
\Tr\,\ln\,{\hat H} = \Tr\,\ln\,
\left(\matrix{
A_{\mu\nu,\al\be}  & 0  \cr
0                  & - 1/2 \cr}
\right)+
\Tr\,\ln\,
\left(\matrix{
\de_{\mu\nu,\al\be}\,{\Box}^3   &   0               \cr
0                               &  {\Box}           \cr} \right)
$$
$$
+\Tr\,\ln\,\left[
\left(\matrix{
{\de_{\al\be,}}^{\mu\nu}  & 0     \cr
0                           & 1     \cr}
\right) +
\left(\matrix{
0  & \left[ P^{\la\tau}\right]_{\mu\nu}\,\na_\la\,\na_\tau\,{\Box}^{-1}
\cr
\left[ Q^{\la\tau}\right]_{\al\be} \,\na_\la\,\na_\tau\,{\Box}^{-3}
&  0    \cr} \right)
\right]                         \eqno(C.4)
$$

The first term in the last expression is the logarithm of the
determinant of a $c$-matrix, and thus it is finite. The second
term depends only on the metric but not on the background scalar,
and therefore it can give only contributions to the renormalization
of the gravitational sector. Since this
renormalization has been already considered in the  diagrams
of Figure 2, we have to ignore here its contributions.  Expanding
the last logarithm   in power series, one can keep only the
terms with proper background dimension. Power counting tells us
that only these terms are divergent (one can also use
universal trace formulae of \cite{bavi} to check the convergence of the
rest of the series). The $Z\,\phi^2$-type divergence is given by
the second term of the series and has the form
$$
\frac{1}{12\,\varepsilon}\;\int d^4x\sqrt{-g}\;
\left[ P^{\la\tau}\right]_{\mu\nu}
A_{\mu\nu,\al\be}
\left[ Q^{\rho\si}\right]_{\al\be}
\,g^{(2)}_{\la\tau\rho\si}                  \eqno(C.5)
$$
Substituting the above
expressions for $A,P,Q$ and taking into account the contribution of
the second diagram of Figure 3, we get the following divergent
part of the effective action
$$
\Ga^{div}_{scalar} = \frac{1}{\varepsilon} \;
\frac{3\,\xi}{\om_1^1+3\,\om_1^2}
\,\left( \frac{3\,\om_1^1 + 10\,\om_1^2}{\om_1^1}
- \xi \right)\,\int d^4x \sqrt{-g}\; \phi^2            \eqno(C.6)
$$

\newpage
\vskip 6mm
{\large \bf FIGURE CAPTIONS}
\vskip 2mm
\noindent {\bf Figure 1:}
Diagrams contributing to the renormalization of
matter and gauge field (thick lines) interactions in a
gravitational classical background (wavy lines).

\vspace*{0.5cm}

\noindent {\bf Figure 2:}
Diagrams contributing to the renormalization of gravitational,
matter and gauge field interactions including quantum
gravity corrections.

\vspace*{0.5cm}

\noindent {\bf Figure 3:}
Divergent diagrams contributing to the mass of scalar particles
in the $N=1$ theory. The divergent contribution is generated by
the non-minimal couplings $${1\over 2} \xi\int d^4x\sqrt{-g}\;
\,\phi^2
$$

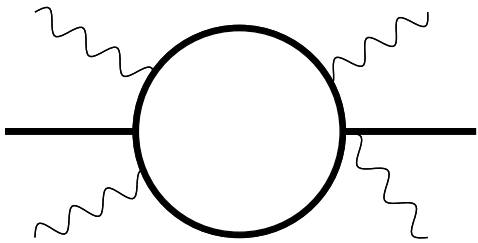
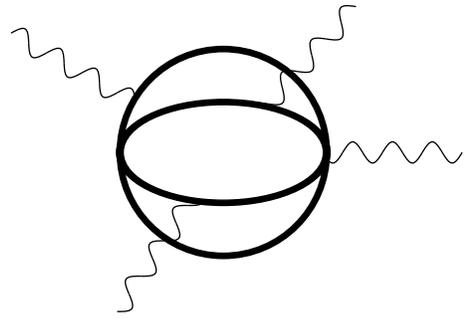
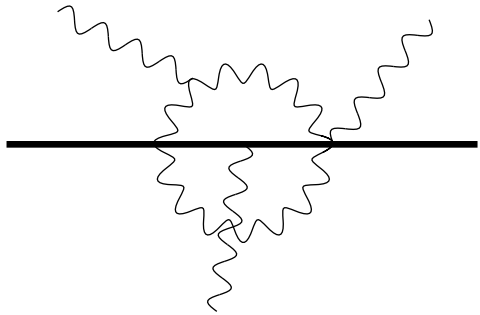
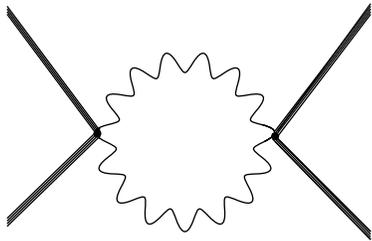
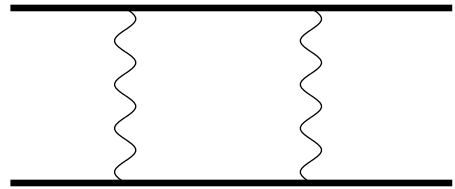
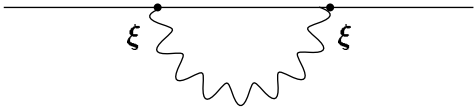
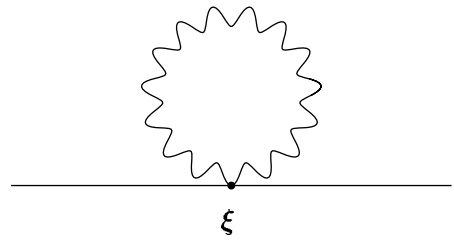
\end{document}